\documentclass[aps,amsmath,prl,reprint,showpacs,preprintnumbers,twocolumn]{revtex4}
\usepackage{amsfonts}
\usepackage{graphicx}
\usepackage{dcolumn}
\usepackage{pdfsync}
\usepackage{multirow}
\usepackage{subfigure}
\usepackage{palatino}
\usepackage{graphicx}
\usepackage{epsfig}
\usepackage{ifpdf}
\ifpdf \DeclareGraphicsExtensions{.pdf,.png,.jpg,.mps}
 \else
\DeclareGraphicsExtensions{.eps,.pdf} \fi

\begin{document}
\title{A frequency determination method for digitized NMR signals}

\author{H.Yan}
\email[Corresponding author: ]{haiyan@umail.iu.edu}
\author{K.Li, R.Khatiwada,E.Smith,W. M. Snow}
\affiliation{Indiana University, Bloomington, Indiana 47408, USA}
\affiliation{Center for Exploration of Energy and Matter, Indiana University, Bloomington, IN 47408}
\author{C.B.Fu}
\affiliation{Department of Physics, Shanghai Jiaotong University,Shanghai,200240,China}
\affiliation{Center for Exploration of Energy and Matter, Indiana University, Bloomington, IN 47408}
\author{P.-H.Chu,H.Gao,W.Zheng}
\affiliation{Triangle Universities Nuclear Laboratory and Department of Physics,
Duke University, Durham, North Carolina 27708, USA}
\date{\today}
\begin{abstract}
We present a high precision frequency determination method for digitized NMR FID signals.  The method employs high precision numerical integration rather than simple summation as in many other techniques. With no independent knowledge of the other parameters of a NMR FID signal (phase $\phi$, amplitude $A$,  and transverse relaxation time $T_{2}$) this method can determine the signal frequency $f_{0}$
with a precision of $1/(8\pi^{2}f_{0}^{2}T_{2}^{2})$ if the observation time $T$ is long enough. The method is especially convenient when the detailed shape of the observed FT NMR spectrum is not well defined. When $T_{2}$ is $+\infty$ and the signal becomes pure sinusoidal, the precision of the method is $3/(2\pi^{2}f_{0}^{2}T^{2})$ which is one order more precise than a typical frequency counter.
Analysis of this method shows that the integration reduces the noise by bandwidth narrowing as in a lock-in amplifier, and no extra signal filters are needed. For a pure sinusoidal signal we find from numerical simulations that the noise-induced error in this method reaches the Cramer-Rao Lower Band(CRLB) on frequency determination. For the damped sinusoidal case of most interest, the noise-induced error is found to be within a factor of 2 of CRLB when the measurement time $T$ is a few times larger than $T_{2}$.We discuss possible improvements for the precision of this method.
\end{abstract}
\maketitle

\section{Introduction}
In nuclear magnetic resonance (NMR) one often encounters a free induction decay (FID) signal $S(t)$ which takes the form of a sinusoidal function multiplied by a decaying exponential:
\begin{equation}\label{eq1}
S(t)=A\cos{(\omega_{0}t+\phi_{0})}\exp{(-\frac{t}{T_{2}})}
\end{equation}
where $t$ is time, $A$ is the signal amplitude, $\omega_{0}=2\pi f_{0}$ is the resonance frequency, $\phi_{0}$ is the signal phase, and $T_{2}$ is the transverse spin relaxation time. In practice, parameters, like $f_{0}$, $\phi_{0}$, etc., cannot be determined without error even in the absence of noise, since $S(t)$ cannot neither be digitized with infinitesimal time intervals nor observed for an infinitely long time. It is of a general interest to determine these parameters using various types of analysis. In particular, the determination of the resonance frequency precisely for a digitized FID signal $S(t)$ observed over a finite time is crucial for recent experiments\cite{ZHE12,BUL12,CHU12} searching for possible new spin dependent interactions which, if present, would cause a tiny shift of the resonance frequency.\\

When $T_{2}=+\infty$, Eqn.(\ref{eq1}) can be simplified to:
\begin{equation}
S(t)=A\cos{(2\pi f_{0}t+\phi_{0})}
\end{equation}
In this case, many different algorithms using Fast Fourier transform(FFT) or Digital Fourier transform (DFT)~\cite{YAN01,TER03} were developed for frequency and spectra estimation in power systems. For sinusoidal signals, by using $\ddot{S}(t)=-\omega_{0}^{2}S(t)$, one can obtain $\omega_{0}$~\cite{OBU07} from a linear fit of $\ddot{S}(t)$ to $S(t)$, where $\ddot{S}(t)$ is derived by finite differentiation of the digitized signal $S(t)$, but extra noise filtering is needed since the second derivative is susceptible to high frequency noise. \\
\begin{figure}[htb]
 \includegraphics[scale=0.5, angle=0]{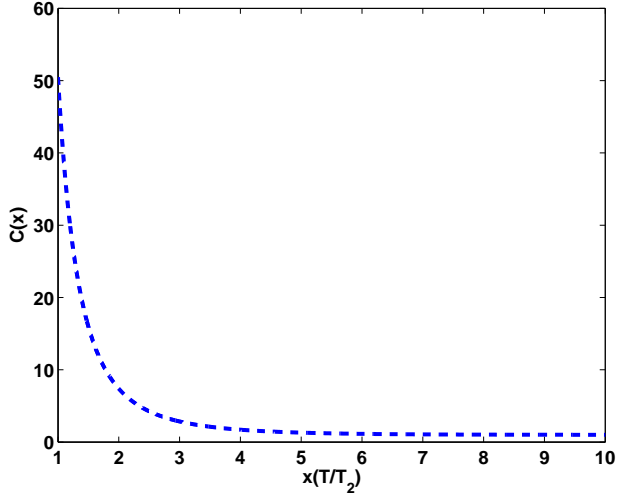}
 \caption{\small{Color online,plot of $C(x)$ of $x$ in range $[1,10]$.}}
 \label{fig.1}
\end{figure}
To determine the frequency precisely and to reduce the noise without filtering, we propose a different approach in this paper based on integration.  We argue that our approach is especially valuable in situations when the shape of the signal in frequency space possesses bias. The structure of this paper is as follows. We first describe the basic principle of the method with an example. We then thoroughly analyze the method and derive its precision. The effect of noise is discussed in the following section. Possible improvements are discussed in the conclusion. \\

\section{The Basic Principle}
Consider a pure sinusoidal signal $S(t)=A\cos{\omega_{0}t}$ observed for a finite time $T$. By multiplying $S(t)$ by another sinusoidal function of frequency $\omega$ and integrating over a time interval of length $T$, a function $\mathcal{L}$ of $\omega$ can be defined as:
\begin{equation}
\mathcal{L}(\omega)=\frac{1}{T}\int^{T}_{0}A\cos{(\omega_{0}t)}\cos{(\omega t)}dt
\end{equation}
If the observing time $T$ is long enough, $\mathcal{L}(\omega)$ will be maximized at $\omega=\omega_{0}$. The frequency determination problem becomes a maximization problem which can be solved by various standard methods, and the remaining problem of the integration with high precision can be addressed using many techniques. For digitized real time signal over an interval $\Delta t$, the error of the integration using a trapezoidal method is $\mathcal{O}(\Delta t/T)^{2}$. An improvement on the integration precision can be achieved by using Richardson's extrapolation strategy~\cite{PRE89}, and the precision of $\mathcal{O}(\Delta t/T)^{4}$ can be obtained by using Simpson's method. Higher precision as $\mathcal{O}(\Delta t/T)^{6}$,$\mathcal{O}(\Delta t/T)^{8}$ can be obtained if necessary by applying the same strategy. In~\cite{CHU12}, $\Delta t\sim10^{-6}s$ and $T\sim10s$. The precision would be $\mathcal{O}(\Delta t/T)^{4}\sim10^{-28}$ so that Simpson's method is accurate enough for our purposes. Therefore from now on in this paper, we ignore  the error caused by numerical integration and assume it is zero. \\

\subsection{Frequency Determination and Precision}
We will next analyze the integration and maximization method for precise frequency determination. If only the frequency is to be determined, the function of $\omega$ can be defined as follows:
\begin{equation}\label{eq.4}
\mathcal{L}_{S}(\omega)=\frac{1}{T^{2}}\{[\int^{T}_{0}S(t)\cos{(\omega t)}dt]^{2}+[\int^{T}_{0}S(t)\sin{(\omega t)}dt]^{2}\}
\end{equation}
or, in the complex notation:
\begin{eqnarray}\label{eq.5}
\mathcal{L}_{S}(\omega)=\frac{1}{T^{2}}|\int^{T}_{0}S(t)\exp{(i\omega t)}dt|^{2}\\
=\frac{A^{2}}{T^{2}}|\int^{T}_{0}\cos{(\omega_{0}t+\phi_{0})}\exp{(-t\Gamma_{2})}\exp{(i\omega t)}dt|^{2}
\end{eqnarray}
where $\Gamma_{2}=1/T_{2}$ is the transverse relaxation time.
A brute force calculation of the above integration gives:
\begin{widetext}
\begin{eqnarray*}
\mathcal{L}_{S}(\omega)=\frac{A^{2}}{4T^{2}}\{\frac{1+\exp{(-2\Gamma_{2} T)}-2\exp{(-\Gamma_{2}T)}\cos{(\omega+\omega_{0})T}}{(\omega+\omega_{0})^{2}+\Gamma_{2}^{2}}
+\frac{1+\exp{(-2\Gamma_{2} T)}-2\exp{(-\Gamma_{2}T)}\cos{(\omega-\omega_{0})T}}{(\omega-\omega_{0})^{2}+\Gamma_{2}^{2}}\\
+\frac{2(\omega^{2}-\omega_{0}^{2}+\Gamma_{2}^{2})\{\cos{2\phi_{0}}[1-\exp{(-2\Gamma_{2} T)}]+4\exp{(-\Gamma_{2}T)}\sin{\frac{(\omega+\omega_{0})T}{2}}\sin{\frac{(\omega-\omega_{0})T}{2}}\cos{(\omega_{0}T-2\phi_{0})}\}}{[(\omega+\omega_{0})^{2}+\Gamma_{2}^{2}][(\omega-\omega_{0})^{2}+\Gamma_{2}^{2}]}\\
\end{eqnarray*}
\begin{eqnarray}\label{eq7}
+\frac{4\Gamma_{2}\omega_{0}\{-\sin{2\phi_{0}}[1-\exp{(-2\Gamma_{2}T)}]
+4\exp{(-\Gamma_{2}T)}\sin{\frac{(\omega+\omega_{0})T}{2}}\sin{\frac{(\omega-\omega_{0})T}{2}}\sin{(\omega_{0}T-2\phi_{0})}\}}{[(\omega+\omega_{0})^{2}+\Gamma_{2}^{2}][(\omega-\omega_{0})^{2}+\Gamma_{2}^{2}]}\}
\end{eqnarray}
\end{widetext}
If we assume that $\omega_{0}$ is large and $T_{2}$ is not too short then $\Gamma_{2}$ is small and the second term on the right hand side of Eqn.(\ref{eq7}) contributes the most to $\mathcal{L}_{S}$.
Or if we let $\omega\rightarrow\omega_{0}$,$\omega_{0}\rightarrow+\infty$ and $\Gamma_{2}\rightarrow0$, only the second term survives which yields $\mathcal{L}_{S}(\omega)\rightarrow A^{2}$. In practice the condition
$\omega_{0}\gg1$ is often satisfied, (in~\cite{CHU12}, for example,  $\omega_{0}=2\pi f_{0}$ and $f_{0}\sim 2.4\times10^{4}Hz$, while $\Gamma_{2}=1/T_{2}$ is not close to $0$, $T_{2}\sim 10 s$). Defining $x=\Gamma_{2}T=T/T_{2}$, we expand $\mathcal{L}_{S}$ around $\omega=\omega_{0}+\delta\omega$ to second order in $\delta\omega$ and $1/\omega_{0}$ to get
\begin{equation}\label{eq.8}
\mathcal{L}_{S}(\omega_{0}+\delta\omega)\approx\frac{A^{2}}{4T^{2}}({a\delta\omega^{2}+b\delta\omega+c})
\end{equation}
where $a,b,c$ are constants depending on $\omega_{0},x,T_{2},\phi$ and $T$:
\begin{widetext}
\begin{eqnarray}
a={T_{2}^{4}}[x^{2}e^{-x}-(1-e^{-x})^{2}]\\
b=2{T_{2}^{2}}[\frac{\cos{2\phi}(1-e^{-2x})}{\omega_{0}}+\frac{Te^{-x}\sin{\omega_{0}T}\sin{(\omega_{0}T-2\phi)}}{\omega_{0}}]\\
c=T_{2}^{2}(1-e^{-x})^{2}+\frac{1+e^{-2x}-2e^{-x}\cos{2\omega_{0}T}}{4\omega_{0}^{2}+1/T_{2}^{2}}+\frac{\cos{2\phi}(1-e^{-2x})}{2\omega_{0}^{2}}-\frac{\sin{2\phi}(1-e^{-2x})}{\omega_{0}/T_{2}}
\approx T_{2}^{2}(1-e^{-x})^{2}
\end{eqnarray}
\end{widetext}

Obviously if $a<0$ and $c>0$,  $b=0$,$\mathcal{L}_{S}(\omega)$ is maximized at $\delta\omega=0$, i.e., $\omega=\omega_{0}$.  However in practice $b$ is usually not $0$ and it will cause a small shift of $\omega$ around $\omega_{0}$:
\begin{widetext}
\begin{equation}
\delta\omega=-\frac{b}{2a}=\frac{\Gamma_{2}^{2}}{\omega_{0}}\frac{\cos{(2\phi_{0})}(1-e^{-2x})+2xe^{-x}\sin{2\omega_{0}T}\sin{(\omega_{0}T-2\phi_{0})}}{(1-e^{-x})^{2}-x^{2}e^{-x}}
\end{equation}
\end{widetext}

It is easy to show that $a<0$ and $c>0$ which proves $L_{S}(\omega)$ is maximized around $\omega\sim\omega_{0}$. Substituting $\omega=2\pi f$ and $\Gamma_{2}=1/T_{2}$, we have:
\begin{widetext}
\begin{equation}\label{eq.10}
\delta f=\frac{1}{8\pi^{2}f_{0}T_{2}^{2}}D(x,\phi_{0})
\end{equation}
\begin{equation}
D(x,\phi_{0})=\frac{\cos{(2\phi_{0})}(1-e^{-2x})+2xe^{-x}\sin{(\omega_{0}T)}\sin{(\omega_{0}T-2\phi_{0})}}{(1-e^{-x})^{2}-x^{2}e^{-x}}
\end{equation}
\end{widetext}

Using
\begin{equation}
|D(x,\phi_{0})|\leq \frac{(1-e^{-2x})+2xe^{-x}}{(1-e^{-x})^{2}-x^{2}e^{-x}}=C(x)
\end{equation}
We obtain
\begin{equation}\label{eq.13}
|\frac{\delta f}{f}|\leq\frac{1}{8\pi^{2}f_{0}^{2}T_{2}^{2}}C(x)
\end{equation}

$C(x)$ is plotted in Figure 1. When $x\rightarrow0+$ and $C(x)\rightarrow+\infty$  the observation time $T$ is very short in comparison to $T_{2}$ and the error for the frequency determination is infinitely large even though $f_{0}$ is large and $T_{2}$ is finite. When $x=T/T_{2}$ exceeds $1$,  $C(x)$ decreases quickly to its asymptotic value of $1$. Once $T_{2}$ and $f_{0}$ are roughly known, the precision of the method can be estimated from $C(x)$. Assuming $T$ is large enough, $x=T/T_{2}$ can be assumed infinite and $C(x)=1$ in this case:
\begin{equation}\label{errD}
|\frac{\delta f}{f}|\sim\frac{1}{8\pi^{2} f_{0}^{2}T_{2}^{2}}
\end{equation}
In deriving Eqn.(\ref{eq.13}) $T_{2}$ is assumed to be finite. $T_{2}$ could be nearly infinite as often encountered in power system applications: in this case by similar reasoning the error for this method is found to be:
\begin{equation}\label{errS}
|\frac{\delta f}{f}|\sim\frac{3|\cos{(\omega_{0}T)}\cos{(\omega_{0}T-2\phi_{0})}|}{2\pi^{2}f_{0}^{2}T^{2}}\leq\frac{3}{2\pi^{2}f_{0}^{2}T^{2}}
\end{equation}
while for this case:
\begin{widetext}
\begin{eqnarray}
a=-\frac{T^{4}}{12}\\
b=T^{2}\frac{\cos{(\omega_{0}T)}\cos{(\omega_{0}T+2\phi)}}{\omega_{0}}\\
c=T^{2}[1+\frac{2\sin{(\omega_{0}T)\cos{(\omega_{0}T-2\phi)}}}{\omega_{0}T}
+\frac{1-\cos{(2\omega_{0}T)}}{2\omega_{0}^{2}T^{2}}]\approx T^{2}
\end{eqnarray}
\end{widetext}

\section{Noise}
 When noise $N(t)$ is included, the function $\mathcal{L}_{SN}(\omega)$ can be defined:
\begin{eqnarray}\label{eq.16}
\mathcal{L}_{SN}(\omega)=\frac{1}{T^{2}}|\int^{T}_{0}[S(t)+N(t)]\exp{(i\omega t)}dt|^{2}
\end{eqnarray}
For the experiments under consideration the signal to noise ratio (SNR) of the final signal is usually $\sim100$, thus the second term of Eqn.(\ref{eq.16}) is much smaller than the first term.
How to estimate $|\int^{T}_{0}N(t)\exp{(i\omega t)}dt|$ is the key to solving the noise problem. According to~\cite{LIB03} we have:
\begin{equation}\label{eq.17}
\frac{1}{T^{2}}|\int^{T}_{0}N(t)\exp{(i\omega t)}dt|^{2}=\sigma^{2}{\frac{B}{f_{BW}}}
\end{equation}
where $\sigma$ is the noise variance, $f_{BW}=1/2\Delta t$ is the sampling rate limited bandwidth, and $B=1/T$ is the measurement bandwidth. As discussed~\cite{LIB03} for the noise the integration according to Eqn.(\ref{eq.17})
 is equivalent to reducing the bandwidth in the frequency domain to $B=1/T$. This is also the principle of how lock-in amplifiers reduce noise, and it is not a surprise that the same noise reduction principle would also work for the algorithm presented in the previous section.

For a sinusoidal signal, according to Ref.\cite{KAY11,GEM10}, the Cramer-Rao lower bound(CRLB) sets the lower limit of the frequency error for any
method as:
\begin{equation}\label{crlb_D}
\delta f_{N}\geq\frac{\sqrt{12/f_{BW}}}{2\pi (A/\sigma) T^{3/2}}
\end{equation}
 For the damped sinusoidal case,
by the standard approach described in~\cite{KAY11} after some manipulation the CRLB bound derived in Ref.\cite{GEM10} can be expressed as:
\begin{eqnarray}\label{crlb_S}
\delta f_{N}=\frac{\sqrt{{1}/f_{BW}}\sqrt{8(1-e^{-2x})}}{2\pi(A/\sigma)T_{2}^{3/2}\sqrt{(1-e^{-2x})^2-4x^2e^{-2x}}}
\end{eqnarray}

 \begin{figure}[htbp]
 \includegraphics[scale=0.55, angle=0]{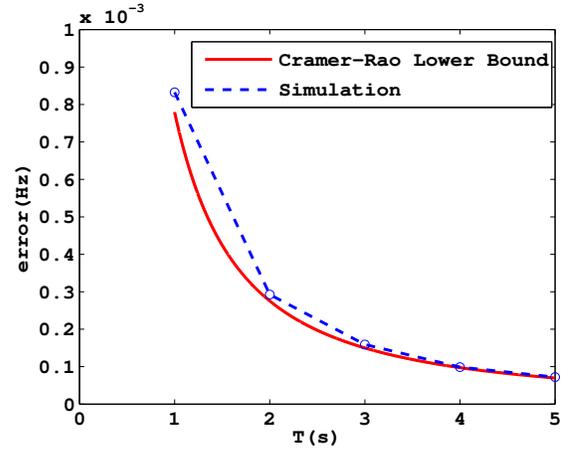}
 \caption{\small{Color online, plot of Cramer-Rao Lower Bound and the error of the presented method obtained by simulation for a sinusoidal signal. The input signal has a frequency of 24000Hz, and white gaussian noise is added
 with SNR=1, and the data sampling rate is $10^{6}$ Hz.}}
 \label{fig.1}
\end{figure}
Since the method can reduce noise by integration, the estimation of the frequency error is expected to be close to the CRLB. Numerical simulations are done to verify this noise reduction characteristic of the method. A signal with known input frequency is first generated then White Gaussian Noise(WGN) is added with a predetermined SNR. The frequency determination method presented is then applied to obtain an output frequency as the approximation of the input.
For the same input signal and measurement time, the same procedure is repeated for many times(1000 for each measurement time), and each time independent WGN is added. Over many trials the standard deviation of the output frequencies gives the error if the
difference between the output mean and input frequency is much smaller. A small SNR(=1) is chosen when generating the noise so that the numerical error is negligible compare to the noise induced error, according to Eqn.(\ref{errD}),(\ref{errS}),(\ref{crlb_D}) and (\ref{crlb_S}). For this small SNR, the method works well. For the pure sinusoidal case, the result
is shown as FIG.2. The errors obtained from simulations match the CRLB very well. For the damped sinusoidal case shown by FIG.3, the errors found by simulation are slightly larger($\sim 15\%$) than CRLB when x$(=T/T_{2})$ is smaller than 3, and increases quickly as x increased to 5. This is not surprising since the actual SNR decreases exponentially as x increases.

 \begin{figure}[htbp]
 \includegraphics[scale=0.55, angle=0]{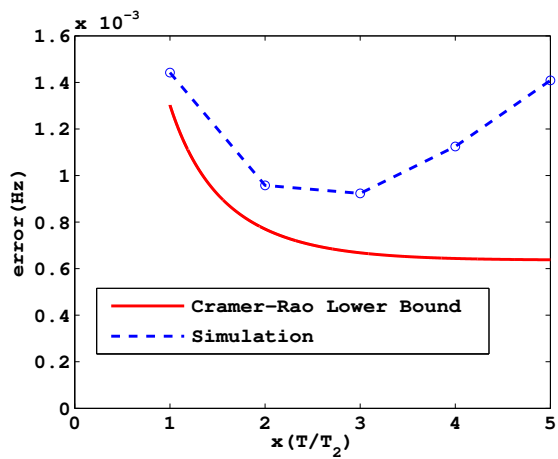}
 \caption{\small{Color online,Color online,plot of Cramer-Rao Lower Bound and the error of the presented method obtained by simulation for a damped sinusoidal signal. The input signal has a frequency of 24000Hz, white gaussian noise is added with SNR=1, $T_{2}=1s$,data sampling rate $10^{6}$ Hz.}}
 \label{fig.1}
\end{figure}

\section{Conclusion and Discussion}
We present to our knowledge a previously unrecognized method for frequency determination for FID NMR signals. The method is based on numerical methods of high precision integration and maximum value location. The same principle works both for sinusoidal and damped sinusoidal signals. For the damped sinusoidal wave, the precision is $1/(8\pi^{2}f_{0}^{2}T_{2}^{2})$ which is limited by $T_{2}$ and the signal frequency if the observation time is long enough. For a pure sinusoidal wave the precision is limited by observation time and signal frequency, and it can be expressed as $3/(2\pi^{2}f_{0}^{2}T^{2})$. The precision is one order of magnitude better than a frequency counter which has precision of $1/f_{0}$~\cite{AGI}. The proposed algorithm is not hard to realize with digital electronics, thus it is possible to build a more precise frequency counter based on this method. We expect that this method will be especially useful in situations when the shape of the Fourier transformed signal is not well defined.\\

It is possible to further improve the precision.  For the pure sinusoidal case once the frequency is obtained the phase factor $\phi_{0}$ can be obtained by locating the maximum from the following integration:

\begin{equation}
\mathcal{L}(\phi)=\frac{1}{T}\int^{T}_{0}A\cos{(\omega_{0}t+\phi_{0})}\cos{(\omega_{0} t+\phi)}dt
\end{equation}
Once the phase and frequency are determined for a pure sinusoidal signal, the frequency shift from the numerical method can be predicted using Eqn.(\ref{errS}) and better precision can be achieved since part of the undesired frequency shift is eliminated. This precision improvement was verified by computer simulations. For the damped signal the same strategy could also work except in this case $T_{2}$ has to be determined precisely by an independent method.\\

When taking into account noise the integration method works as a narrow bandwidth filter around the signal frequency like a lock-in amplifier. For pure sinusoidal signal the estimated error caused by noise is found to reach the Cramer-Rao Lower Bound~\cite{KAY11}. For the damped sinusoidal case the estimated error is within factor of 2 of the CRLB derived in literature when x($=T/T_{2}$) is less than 5. The method works well for strong white noise case as SNR=1.

\section{Acknowledgements}

This work was supported by the Department of Energy and by NSF grant PHY-1068712. W. M. Snow, H, Yan, K.Li, R.Khatiwada, and E.Smith acknowledge support from the Indiana University Center for Spacetime Symmetries and the Indiana University Collaborative Research Grant program.

\end{document}